\documentclass[copyright,creativecommons]{eptcs}

\usepackage{breakurl, 
amssymb, 
amsmath, 
amsthm, 
}
\providecommand{\event}{CCA 2010}
\newtheorem{defn}{Definition}[section]

\newtheorem{theo}[defn]{Theorem}

\newtheorem{prop}[defn]{Proposition}

\newtheorem{lemm}[defn]{Lemma}

\newtheorem{coro}[defn]{Corollary}

\newenvironment{defi}{\begin{defn} \rm}
                     {\end{defn}}

\newtheorem*{prov}{Proof}

\newtheorem*{skpr}{Sketch of Proof}

\newtheorem*{remark}{Remark}

\newenvironment{prove}{\begin{prov} \rm}
                     {\end{prov}}

\newenvironment{prsk}{\begin{skpr} \rm}
                     {\end{skpr}}

\newenvironment{rem}{\begin{remark} \rm}
                     {\end{remark}}

\newcommand{\cB}{\mathcal{B}}
\newcommand{\cC}{\mathcal{C}}
\newcommand{\cS}{\mathcal{S}}
\newcommand{\cO}{\mathcal{O}}

\newcommand{\cG}{\mathcal{G}}

\newcommand{\Q}{\mathbb{Q}}

\newcommand{\id}{\mathrm{id}}

\newcommand{\ifof}{\Leftrightarrow}

\newcommand{\dsup}{\bigvee ^\uparrow}

\newcommand{\w}{\omega}

\newcommand{\Sub}{\mathrm{Sub}_\sigma}

\newcommand{\Eq}{\mathrm{Eq}}
\newcommand{\Sm}{\mathrm{Sm}}

\newcommand{\qedd}{\nobreak \ifvmode \relax \else

           \ifdim\lastskip<1.5em \hskip-\lastskip

           \hskip1.5em plus0em minus0.5em \fi \nobreak

           \vrule height0.75em width0.5em depth0.25em\fi}

\title{A domain-theoretic investigation of posets of sub-$\sigma$-algebras\\\begin{small}(Extended Abstract)                                                                                               \end{small}}
\author{Ingo Battenfeld
\institute{Fakult\"at f\"ur Informatik\\TU Dortmund, Germany}
\email{battenfeld@ls10.cs.tu-dortmund.de}
}
\def\titlerunning{A domain-theoretic investigation of posets of sub-$\sigma$-algebras}
\def\authorrunning{Ingo Battenfeld}
\begin{document}

\maketitle

\begin{abstract}
Given a measurable space $(X, \mathcal{M})$ there is a (Galois) connection between sub-$\sigma$-algebras of $\mathcal{M}$ and equivalence relations on $X$. On the other hand equivalence relations on $X$ are closely related to congruences on stochastic relations. In recent work, Doberkat has examined lattice properties of posets of congruences on a stochastic relation and motivated a domain-theoretic investigation of these ordered sets. Here we show that the posets of sub-$\sigma$-algebras of a measurable space do not enjoy desired domain-theoretic properties and that our counterexamples can be applied to the set of smooth equivalence relations on an analytic space, thus giving a rather unsatisfactory answer to Doberkat's question.
\end{abstract}

\section{Introduction}

Recently, measurable spaces, and in particular analytic spaces, have been proposed for modelling probabilistic systems \cite{deshetalbisim,eedcongbisim,dadelapa}. The continuous, uncountable nature of such systems is reflected by the uncountable state spaces provided by the measurable setting. Many notions of equivalence or similarity of probabilistic systems, such as behavioural equivalence, logical equivalence or bisimulation, can be modelled in this setting and can then be used for system simplification and system verification. One of the tools in modelling equivalences and bisimilarity in the setting of analytic spaces is the notion of a congruence on a stochastic relation \cite{eed1,eedstochrel}. Such a congruence is given by a pair of smooth equivalence relations on the underlying analytic spaces satisfying certain conditions.

In recent work, Doberkat has investigated the set of smooth equivalence relations of an analytic space on order- and lattice-theoretic properties. Results include an example showing that congruences are not closed under intersection in the lattice of all equivalence relations \cite{eedconglatt} and a construction having similar properties to a complement or negation \cite{eednegcong}. In the end of this second article he asks for a domain-theoretic investigation of the set of smooth equivalence relation on an analytic space, more specifically for a characterisation of the way-below order on posets of smooth relations.

The way-below order comes from domain theory \cite{abjudomain}, probably the most popular framework for denotational semantics of programming languages. The key idea of domain theory is to equip a set denoting a datatype with an order which reflects the amount of information that the respective data carries. The way-below order is derived from this original order to model uniform approximation of data elements and programs. Thus, a well-behaved way-below order on smooth equivalence relations may open the door towards a theory of approximations and computability of certain kind of similarities.

In this paper we attempt a domain-theoretic investigation of posets of sub-$\sigma$-algebras. During this investigation, we generalise one of Doberkat's results in \cite{eedconglatt} and show that the posets of smooth equivalence relations on an analytic space ordered by inclusion and reverse inclusion might have trivial orders of approximation, hence they do not form pleasant domains in general. To obtain this result, we will use the isomorphism between smooth equivalence relations on the analytic space $(X, \mathcal{M})$ and countably-generated sub-$\sigma$-algebras of $\mathcal{M}$, as defined in section 1.7 of \cite{eed2}. This isomorphism is induced by a Galois connection between the set of all equivalence relations on $X$ and sub-$\sigma$-algebras of $\mathcal{M}$, as mentioned in \cite{dadelapa}. We also use results of \cite{raorao}, where the general lattice-theoretic properties of the set of sub-$\sigma$-algebras of a given $\sigma$-algebra $\mathcal{M}$ are examined.

The paper is organised as follows. In section 2, we fix our measure theoretic and domain theoretic notation and recall some basic results, in particular the Galois connection between sub-$\sigma$-algebras and equivalence relations. In section 3, we investigate posets of sub-$\sigma$-algebras of a given $\sigma$-algebra $\mathcal{M}$ with respect to the set-inclusion order. Along the way we show that the countably-generated sub-$\sigma$-algebras of an analytic space do in general not form directed complete posets and use this result to prove that the category of analytic spaces does not have pushouts. Finally, we show that with respect to the inclusion order, various posets of sub-$\sigma$-algebras of the usual Borel-$\sigma$-algebra of the real numbers have a trivial approximation order. In section 4, we investigate the opposite of the inclusion order on the posets of sub-$\sigma$-algebras, which arguably is the more interesting order from the category-theoretical viewpoint. Under the Galois connection it corresponds directly to the inclusion order on the equivalence relations. We show that also with respect to this order the posets of sub-$\sigma$-algebras are not directed complete and might have a trivial way-below order. Again we show that this in particular holds for the standard Borel $\sigma$-algebra on the real numbers and hence the set of smooth relations on this measurable space has a trivial order of approximation when equipped with the usual inclusion order. We finish the paper with some final remarks in section 5.

\section{Preliminaries}

Let us start by introducing our notation and giving some basic results in this section. First we give the measure-theoretic background, followed by equivalence relations on measurable spaces and the Galois connection between them and sub-$\sigma$-algebras, and finally the domain-theoretic counterpart.

\subsection{Measurable spaces}

We do not give details or motivations about the basic notions in this section, for a more detailed treatment the reader is referred to \cite{eed1,srivastava}.

\begin{itemize}
\item A \emph{measurable space} $(X, \mathcal{M})$ is a set $X$ equipped with a $\sigma$-algebra $\mathcal{M}$ of subsets of $X$. The elements of $\mathcal{M}$ are called \emph{measurable sets}. A map $f: (X, \mathcal{M}) \rightarrow (Y, \mathcal{N})$ between measurable spaces is called a \emph{measurable map} if for all $A \in \mathcal{N}$, it holds that $f^{-1} (A) \in \mathcal{M}$.

\item If $\mathcal{M}$ and $\mathcal{N}$ are $\sigma$-algebras on a set $X$ with $\mathcal{N} \subseteq \mathcal{M}$, then $\mathcal{N}$ is a \emph{sub-$\sigma$-algebra} of $\mathcal{M}$.

\item A $\sigma$-algebra $\mathcal{M}$ on $X$ is called \emph{countably-generated} if there exists a sequence $(A_n)_{n \in \mathbb{N}}$ of subsets of $X$ such that $\mathcal{M} = \sigma (\{A_n\}_{n \in \mathbb{N}})$, i.e. the smallest $\Sigma$-algebra containing all the $A_n$.

\item A measurable space $(X, \mathcal{M})$ is called a \emph{Standard Borel space} if there exists topology $\mathcal{T}$ on $X$ such that $(X, \mathcal{T})$ is a Polish space \cite{kuratowski} and $\mathcal{M} = \sigma (\mathcal{T})$, i.e. the smallest $\sigma$-algebra containing $\mathcal{T}$.

\item An \emph{analytic space} $(X, \mathcal{M})$ is a measurable space which is isomorphic to the image of some measurable map $f: (Y, \mathcal{N}) \rightarrow (Y', \mathcal{N}')$ between Standard Borel spaces \cite{kechriscdst}.
\end{itemize}

The topological origin of analytic and Standard Borel spaces yields the machinery to show many interesting results and hence makes them an ideal setting to study continuous probabilistic systems. Some of these results can be extended to the class of all measurable spaces, as done by Schubert in \cite{schubcoalgmodlog}. In this generalisation the following properties, which are always satisfied in analytic spaces, play important r\^{o}les.

\begin{itemize}
\item A measurable space $(X, \mathcal{M})$ is called \emph{separated} if the elements of $\mathcal{M}$ separate the points of $X$, i.e. if for any distinct $x,y \in X$ there exists some $A \in \mathcal{M}$ containing $x$ but not $y$. A measurable space $(X, \mathcal{M})$ is called \emph{separable} if it is separated and $\mathcal{M}$ is countably-generated.
\end{itemize}\label{ref0}

In this paper we are interested in studying the properties of partially ordered sets (henceforth \emph{posets}) of sub-$\sigma$-algebras of measurable and analytic spaces. Notice that for any set $X$ the power set $\mathcal{P}(X)$ forms the largest and $\{\emptyset, X\}$ the smallest $\sigma$-algebra on $X$. Moreover if $\{\mathcal{A}_i\}_{i \in I}$ is a family of $\sigma$-algebras on $X$, then $\bigcap_{i \in I} \mathcal{A}_i$ is also a $\sigma$-algebra on $X$, and it is the largest common sub-$\sigma$-algebra of all the $\mathcal{A}_i$. It follows that the $\sigma$-algebras on a set $X$ form a complete lattice under the usual set-inclusion order. The meet is given by intersection and the join is given as $\bigvee_{i \in I} \mathcal{A}_i = \sigma(\bigcup_{i \in I} \mathcal{A}_i)$.

For a given measurable space $(X, \mathcal{M})$ we fix the following notation.

\begin{itemize}
\item By $\mathrm{Sub}_\sigma(\mathcal{M})$ we denote the set of all sub-$\sigma$-algebras of $\mathcal{M}$. The inclusion order $\subseteq$ makes $(\mathrm{Sub}_\sigma(\mathcal{M}), \subseteq)$ into a complete lattice. We write $\sqsubseteq$ for the opposite order (i.e. $\mathcal{A} \sqsubseteq \mathcal{A}'$ if $\mathcal{A} \supseteq \mathcal{A}'$) on these sets, and then also $(\mathrm{Sub}_\sigma(\mathcal{M}), \sqsubseteq)$ is a complete lattice.
\end{itemize}

In the study of analytic spaces, the countably generated $\sigma$-algebras are of particular interest \cite{eed1,eed2}, motivating the following notation.

\begin{itemize}
\item By $\mathrm{Sub}^{\omega}_\sigma(\mathcal{M})$ we denote the set of countably-generated sub-$\sigma$-algebras of $\mathcal{M}$. With the inherited orders, $(\mathrm{Sub}^{\omega}_\sigma(\mathcal{M}), \subseteq)$ and $(\mathrm{Sub}^{\omega}_\sigma(\mathcal{M}), \sqsubseteq)$ are posets.
\end{itemize}

Notice that the set of all $\sigma$-algebras on a set $X$ is given by $\mathrm{Sub}_\sigma(\mathcal{P}(X))$, and the set of all countably-generated $\sigma$-algebras by $\mathrm{Sub}^{\omega}_\sigma(\mathcal{P}(X))$. Thus our results below can also be adjusted to a setting where no $\sigma$-algebra of measurable sets is assumed.

Let us make the following trivial observation.

\begin{prop}
Let $(X, \mathcal{M})$ be a measurable space. Then $(\mathrm{Sub}^{\omega}_\sigma(\mathcal{M}), \subseteq)$ is a $\vee$-semilattice and has countable suprema, accordingly $(\mathrm{Sub}^{\omega}_\sigma(\mathcal{M}), \sqsubseteq)$ is a $\wedge$-semilattice and has countable infima.\label{refx3}
\end{prop}

Below we will show that in general they do not support the corresponding opposite operation.

\subsection{Equivalence relations}

As we have mentioned in the introduction, there is a close connection between equivalence relations and sub-$\sigma$-algebras of a measurable space $(X, \mathcal{M})$. This connection is made precise by the following result, the proof of which is straightforward.

\begin{lemm}
Let $(X, \mathcal{M})$ be a measurable space and $\sim \subseteq X \times X$ be an equivalence relation. Then the set $\mathcal{M}_\sim$ of $\sim$-invariant sets of $\mathcal{M}$ forms a sub-$\sigma$-algebra of $\mathcal{M}$.
\end{lemm}

Notice that in general not all $\sigma$-algebras are defined like this. For instance consider $\cC$, the $\sigma$-algebra of countable and cocountable subsets of a measurable space $(X, \mathcal{P}(X))$, where $X$ is some uncountable set. Since $\cC$ contains all the singletons, the only equivalence relation which could possibly induce it according to the lemma above is equality. However every subset of $X$ is $=$-invariant, and so we have $\mathcal{P}(X)_= = \mathcal{P}(X) \neq \cC$. Thus, ingeneral, the set of $\sigma$-algebras of invariant sets of equivalence relations forms a proper subset of the set of all sub-$\sigma$-algebras of a measurable space $(X, \mathcal{M})$.

\begin{defi}
Let $(X, \mathcal{M})$ be a measurable space. We denote the set of equivalence relations on $X$ by $\mathrm{Eq} (X)$. For an element $\sim \in \mathrm{Eq} (X)$, the \emph{$\sim$-induced sub-$\sigma$-algebra} $\mathcal{M}_\sim$ of $\mathcal{M}$ is given by the $\sim$-invariant sets of $\mathcal{M}$. We denote the subset of $\Sub (\mathcal{M})$ given by equivalence relation induced sub-$\sigma$-algebras by $\mathrm{Sub}^{\mathrm{eq}}_\sigma (\mathcal{M})$.\label{x0}
\end{defi}

Notice that there is a converse to the construction of $\sigma$-algebras of invariant sets.

\begin{defi}
For a given $\sigma$-algebra $\mathcal{M}$ on a set $X$, we denote by $\sim_{\mathcal{M}}$ the \emph{equivalence relation induced by $\mathcal{M}$}, given by
$$x \: \sim_\mathcal{M} \: y := (\forall A \in \mathcal{M}. \: x \in A \ifof y \in A).$$\label{x1}
\end{defi}

It is well-known that for such equivalence relations it suffices to consider the behaviour on a generator of the given $\sigma$-algebra, as the following result shows, whose proof can be found as Lemma 3.1.6 in \cite{srivastava}.

\begin{lemm}
Let $\cG$ be a generator of a $\sigma$-algebra $\mathcal{M}$ on $X$. Then $x \: \sim_\mathcal{M} \: y$ if and only if for all $G \in \cG$ it holds that:
$$x \in G \; \ifof \; y \in G.$$
\end{lemm}

Of particular interest in the study of analytic spaces have been those equivalence relations which are induced by a countably-generated $\sigma$-algebra.

\begin{defi}
A \emph{smooth relation} on a measurable space $(X, \mathcal{M})$ is an equivalence relation induced by a countably generated sub-$\sigma$-algebra of $\mathcal{M}$. We denote the subset of smooth relations of $\mathrm{Eq}(X)$ by $\mathrm{Sm}(X, \mathcal{M})$.
\end{defi}

Smooth relations are of paramount interest in the study of analytic spaces because they characterise exactly those equivalence relations for which the quotient space is again analytic as the following pivotal result shows. It is shown as Lemma 1.52 and Proposition 1.53 in \cite{eed1}.

\begin{theo}
Let $(X, \mathcal{M})$ be an analytic space and $\sim$ be an equivalence relation on $X$. Then the following are equivalent:
\begin{itemize}
 \item[(I)] $\sim$ is smooth,

 \item[(II)] $\sim$ is the kernel relation for a measurable map $f: (X, \mathcal{M}) \rightarrow (Y, \mathcal{N})$ into an analytic space $(Y, \mathcal{N})$,

 \item[(III)] the quotient space $(X / \! \sim, \mathcal{M}_\sim)$ is analytic.
\end{itemize}\label{x3}
\end{theo}

\begin{coro}
For an analytic space $(X, \mathcal{M})$ the sets $\mathrm{Sub}^{\omega}_\sigma(\mathcal{M})$ and $\mathrm{Sm}(X, \mathcal{M})$ are isomorphic.\label{x2}
\end{coro}

This isomorphism can be extended along intrinsic orders on these sets. In fact, it is the restriction of a more general order theoretic connection between equivalence relations and sub-$\sigma$-algebras on measurable spaces, namely a Galois connection.

\begin{defi}
A \emph{Galois connection} (or \emph{adjunction}) between posets $(X, \leq)$ and $(Y, \sqsubseteq)$ is a pair of monotone maps $f: (X, \leq) \leftrightarrow (Y, \sqsubseteq) :g$ for which it holds that $f(x) \sqsubseteq y$ if and only if $x \leq g(y)$. In this case $f$ is called the \emph{lower adjoint} and $g$ the \emph{upper adjoint}.
\end{defi}

For the remainder of this part, let $(X, \mathcal{M})$ be some fixed measurable space. To avoid confusion, we write $\lessdot$ for the inclusion order on $\Eq (X)$, i.e. $\sim \lessdot \smile$ if $x \sim y$ implies $x \smile y$ for all $x,y \in X$. Then $(\Eq (X), \lessdot)$ is a complete lattice where meet is defined by intersection of the given relations in $X \times X$ and join by the equivalence relation generated by the union of the given relations in $X \times X$. It is essentially shown in \cite{dadelapa} that the constructions defined in Definitions \ref{x0} and \ref{x1} form a Galois connection between $(\mathrm{Eq}(X), \lessdot)$ and $(\mathrm{Sub}_\sigma(\mathcal{M}), \sqsubseteq)$.

\begin{lemm}
For elements $\mathcal{A}, \cB \in \mathrm{Sub}_\sigma(\mathcal{M})$ with $\mathcal{A} \sqsubseteq \cB$, it holds that $\sim_\mathcal{A} \lessdot \sim_\cB$, and conversely for elements $\sim, \smile \in \mathrm{Eq}(\mathcal{M})$ with $\sim \lessdot \smile$, it holds that $\mathcal{A}_\sim \sqsubseteq \mathcal{A}_\smile$. The maps $\mathcal{M}_{(-)} : (\mathrm{Eq}(X), \lessdot)  \leftrightarrow (\Sub (\mathcal{M}), \sqsubseteq) : \sim_{(-)}$ form a Galois connection, with $\mathcal{M}_{(-)}$ being the lower and $\sim_{(-)}$ being the upper adjoint.
\end{lemm}

Notice that in general we do not have an isomorphism between $(\mathrm{Sub}_\sigma(\mathcal{M}), \sqsubseteq)$ and $(\mathrm{Eq}(X), \lessdot)$. The example of the countable-cocountable $\sigma$-algebra above shows that $\mathcal{M}_{(-)}$ is not surjective, and the following consideration, which is Example 4.12 in \cite{dadelapa}, shows that $\sim_{(-)}$ cannot be surjective. Let $T$ be a non-measurable subset of $(\mathbb{R}, \cB)$ (with the usual Euclidean Borel algebra) and define $\sim$ to have equivalence classes $T$ and $\mathbb{R} \setminus T$. Then $\sim$ cannot be induced by any sub-$\sigma$-algebra of $\cB$ as this would have to contain $T$.

Notice that we can characterise the image of one of the two compositions.

\begin{lemm}
The image of $\mathcal{M}_{\sim_{(-)}}$ is given by $\mathrm{Sub}^{\mathrm{eq}}_\sigma (\mathcal{M})$ which therefore is the set of fixed points of $\mathcal{M}_{\sim_{(-)}}$. Thus $(\mathrm{Sub}^{\mathrm{eq}}_\sigma (\mathcal{M}), \sqsubseteq)$ is a complete lattice with join defined as:
$$\bigvee \mathcal{A}_i := \mathcal{M}_{\bigvee \sim_{\mathcal{A}_i}} \hspace{1cm}.$$\label{ref4}
\end{lemm}

The other composition is the more problematic one, and we do not know how to characterise it. Nevertheless we mention Corollary 4.10 of \cite{dadelapa}.

\begin{prop}
Let $(X, \mathcal{M})$ be a measurable space and $\sim$ be an equivalence relation on $X$ such that all equivalence classes are in $\mathcal{M}$. Then $\sim = \sim_{\mathcal{M}_\sim}$.\label{ref3}
\end{prop}

Notice that this is the case for smooth equivalence relations, because we have:

\begin{prop}
For a smooth equivalence relation on a measurable space $(X, \mathcal{M})$ all equivalence classes are measurable sets.\label{x4}
\end{prop}

\begin{prove}
Let $x \in X$ and $(A_n)_{n \in \mathbb{N}}$ be a countable generator inducing the equivalence relation $\sim$. Then it holds that $[x]_\sim = \bigcap \{ A_n |\: x \in A_n\} \cap \bigcap \{ X \setminus A_n |\: x \notin A_n\}$, which is a measurable set.\qed
\end{prove}

Thus, we get that smooth equivalence relations are fixed points of $\sim_{\mathcal{M}_{(-)}}$. As mentioned, on analytic spaces the isomorphism of Corollary \ref{x2} becomes an order isomorphism between $(\Sm (X, \mathcal{M}), \lessdot)$ and $(\mathrm{Sub}^{\omega}_\sigma (\mathcal{M}), \sqsubseteq)$, see e.g. section 1.7 of \cite{eed2}.

\begin{lemm}
On an analytic space $(X, \mathcal{M})$, the Galois connection $(\Sub (\mathcal{M}), \sqsubseteq) \leftrightarrow (\mathrm{Eq}(X), \lessdot)$ restricts to an isomorphism $(\mathrm{Sub}^{\omega}_\sigma (\mathcal{M}), \sqsubseteq) \cong (\mathrm{Sm}(X, \mathcal{M}), \lessdot)$.\label{ref2}
\end{lemm}

Observe, that this result also shows that for an analytic space $(X, \mathcal{M})$ we have $\mathrm{Sub}^{\omega}_\sigma (\mathcal{M}) \subseteq \mathrm{Sub}^{\mathrm{eq}}_\sigma (\mathcal{M}) \subseteq \Sub (\mathcal{M})$. Both inclusions are proper in general. For the last one, this is clear, as the example of the countable-cocountable sub-$\sigma$-algebra on the real numbers has shown. The first one can be obtained by using the counterexample of Section 3.2 in \cite{eedconglatt}. Once we leave the realm of analytic spaces, the first inclusion need not be valid: The usual Euclidean Borel-$\sigma$-algebra on the real numbers is countably generated but not in $\mathrm{Sub}^{\mathrm{eq}}_\sigma(\mathcal{P}(\mathbb{R}))$.

\subsection{Domain Theory}

Also in this part we refrain from giving a motivation or detailed background and simply introduce the concepts we need below. For more details on domain theory we suggest the introductory survey \cite{abjudomain}.

\begin{itemize}
\item Let $(X,\leq)$ be a partially ordered set. A subset $D \subseteq X$ is \emph{directed} if for any $x,y \in D$ there exists some $z \in D$ with $x,y \leq z$. The poset $(X, \leq)$ is called \emph{directed complete} if every directed subset of $X$ has a supremum $\dsup D$ in $X$.

\item In a poset $(X, \leq)$ an element $x$ is \emph{way-below} another element $y$, written as $x \ll y$, if for any directed $D \subseteq X$ which has a supremum $\dsup D$ with $y \leq \dsup D$, there exists some $d \in D$ with $x \leq d$.
\end{itemize}

The way-below relation is of paramount interest in domain theory as it provides a meaning of uniform approximation of elements. It is sometimes called the \emph{order of approximation}.

In fact, directed completeness is a generalisation of chain completeness which we will also use below.

\begin{itemize}
\item Let $(X,\leq)$ be a partially ordered set. By an \emph{$\w$-chain} we refer to a sequence $\{x_n\}_{n \in \mathbb{N}}$ which is increasing in $X$, i.e. for all $n \in \mathbb{N}$, it holds that $x_n \leq x_{n+1}$. The poset $(X, \leq)$ is called \emph{$\w$-chain complete} if every $\w$-chain in $X$ has a supremum $\dsup_n x_n$ in $X$.

\item In a poset $(X, \leq)$ an element $x$ is \emph{$\w$-way-below} another element $y$, written as $x \ll_\w y$, if for any $\w$-chain $\{x_n\}_{n \in \mathbb{N}}$ in $X$ which has a supremum $\dsup_n x_n$ with $y \leq \dsup_n x_n$, there exists some $n_0 \in \mathbb{N}$ with $x \leq x_{n_0}$.
\end{itemize}

These definitions should suffice to follow the technical development in this paper. On a detailed analysis on how the directed and $\w$-chain notions relate we refer the reader to section 2.2.4 of \cite{abjudomain}.

\section{The inclusion order}

In this section we investigate the lattices $(\mathrm{Sub}_\sigma(\mathcal{M}), \subseteq)$ and $(\mathrm{Sub}^{\mathrm{eq}}_\sigma (\mathcal{M}), \subseteq)$ and also the poset $(\Sub^{\w} (\mathcal{M}), \subseteq)$ on their properties. Let us start by recalling some more or less obvious results.

\begin{prop}
For any measurable space $(X, \mathcal{M})$, every element $\mathcal{A} \in \Sub (\mathcal{M})$ can be obtained as the directed supremum of finitely-generated sub-$\sigma$-algebras of $\mathcal{M}$.
\end{prop}

Notice that (surprisingly) the finitely-generated sub-$\sigma$-algebras of $\mathcal{M}$ need not be compact elements in the lattice: Consider a point $x \in \mathbb{R}$ and the $\sigma$-algebras $\mathcal{A}_n := \sigma (\{U_{\frac{1}{k}} (x) |\: k \leq n \})$, where $U_{\frac{1}{k}} (x) := \{y \in X | \: |x-y| < \frac{1}{k} \}$ denotes the open ball around $x$ with radius $\frac{1}{k}$. Then $\{x\} \in \dsup \mathcal{A}_n$ but $\{x \} \notin \mathcal{A}_n$ for any $n \in \mathbb{N}$.

Below we will show below that the way-below order on $(\mathrm{Sub}_\sigma(\cB), \subseteq)$, where $\cB$ is the usual Borel $\sigma$-algebra on $\mathbb{R}$ is trivial. But let us first mention some more general results.

\begin{coro}
Suppose $(\Sub^{\w} (\mathcal{M}), \subseteq)$ is directed complete. Then it is a monotone retract of $(\Sub (\mathcal{M}), \subseteq)$ with section-retraction pair $\iota: (\Sub^{\w} (\mathcal{M}), \subseteq) \leftrightarrow (\Sub (\mathcal{M}), \subseteq): \kappa$ such that $\id_{\Sub (\mathcal{M})} \leq \iota \circ \kappa$. It follows that in this case $\mathcal{M}$ is countably-generated.
\end{coro}

Notice that we do not know whether $\Sub^{\w} (\mathcal{M})$ forms a continuous retract of $\Sub (\mathcal{M})$ if it is directed complete, since the inclusion map $\iota$ might not preserve directed suprema. In fact it will do so only if $\mathrm{Sub}^{\omega}_\sigma (\mathcal{M}) \equiv \Sub (\mathcal{M})$.

Recall a counterexample in section 3.2 of \cite{eedconglatt} which shows that $\Sub^{\w} (\mathcal{M})$ is in general not closed under the intersection operation $\cap$ in $\Sub (\mathcal{M})$ even for analytic spaces.

\begin{prop}
In general, for an analytic space $(X, \mathcal{M})$ the poset $(\Sub^{\w} (\mathcal{M}), \subseteq)$ is not a lattice.
\end{prop}

\begin{prove}
Consider the mentioned example which constructs countably-generated sub-$\sigma$-algebras $\mathcal{A}, \cB$ of an analytic space $(X, \mathcal{M})$, for which $\mathcal{A} \cap \cB$ is not countably-generated. Suppose that nevertheless $(\Sub^{\w} (\mathcal{M}), \subseteq)$ has a meet operation $\wedge^\w$. Then, since $\mathcal{A} \wedge^\w \cB \subseteq \mathcal{A}, \cB$, it holds that $\mathcal{A} \wedge^\w \cB \subseteq \mathcal{A} \cap \cB$ in $\Sub (\mathcal{M})$. Moreover, for each $A \in \mathcal{A} \cap \cB$, it holds that $\sigma (\{A\}) \in \Sub^{\w} (\mathcal{M})$ and $\sigma (\{A\}) \subseteq \mathcal{A}, \cB$. Thus we get that $A \in \sigma (\{A\}) \subseteq \mathcal{A} \wedge^\w \cB$, showing $\mathcal{A} \cap \cB \subseteq \mathcal{A} \wedge^\w \cB$, and we conclude that they must be equal, which contradicts the counterexample.\qed
\end{prove}

This implies the following result, which generalises Doberkat's Corollary 3.19 in \cite{eedconglatt} which shows that analytic spaces are not closed under pushouts in the category of measurable spaces and measurable maps.

\begin{theo}
The category of analytic spaces and measurable maps does not have all pushouts.
\end{theo}

Now we show that the way-below relation on the sub-$\sigma$-algebras $(\Sub (\cB), \subseteq)$ of the standard Borel $\sigma$-algebra $\cB$ on the real numbers $\mathbb{R}$ is trivial.

\begin{theo}
In $(\Sub (\cB), \subseteq)$, it holds that $\mathcal{A} \ll \cB$ if and only if $\mathcal{A} = \{\emptyset, \mathbb{R} \}$.
\end{theo}

\begin{prsk}
Suppose $\{\emptyset, \mathbb{R} \} \neq \mathcal{A}$ and fix a Borel set $B \subseteq \mathbb{R}$ in $\mathcal{A}$ such that $\emptyset \neq B \neq \mathbb{R}$. Pick some point $y \in \mathbb{R}$ of the topological boundary of $\mathcal{A}$ and modify the sets of the generator $\cO := \{ U_q (s) |\: r,s \in \Q \}$ of $\cB$ as follows. Enumerate the sets, e.g. $\{V_n\}_{n \in \mathbb{N}}$, and either add a small ball $U_{\frac{1}{n}} (y)$ to $V_n$ or remove it according as to whether $y$ is a member of $V_n$ or not. The modified sets still generate $\cB$ hence one can obtain $\cB$ as the directed supremum of the $\sigma$-algebras generated by finitely many such sets. However, the modification ensures that none of these finite $\sigma$-algebras contains $B$ and hence cannot have $\mathcal{A}$ as a subset.\qed
\end{prsk}

\begin{coro}
For the way-below relation $\ll$, resp. the $\w$-chain way-below relation $\ll_{\w}$, on $(\Sub (\cB), \subseteq)$, $(\mathrm{Sub}^{\mathrm{eq}}_\sigma (\cB), \subseteq)$ and $(\mathrm{Sub}^{\omega}_\sigma (\cB), \subseteq)$, it holds that $\mathcal{A} \ll \cB$ (resp. $\mathcal{A} \ll_{\w} \cB$) if and only if $\mathcal{A} = \{\emptyset, \mathbb{R}\}$.
\end{coro}

This shows that from a domain-theoretical viewpoint, the inclusion order $\subseteq$ does in general not yield pleasant posets of sub-$\sigma$-algebras, even in the restricted setting of analytic spaces or Standard Borel spaces.

\section{The opposite order}

After our investigation of the inclusion order in the previous section, we now investigate the opposite order on posets of sub-$\sigma$-algebras. Observe that from a category-theoretical viewpoint this order is the more interesting one, since whenever $\mathcal{A} \sqsubseteq \cB$ for $\mathcal{A},\cB \in \Sub (\mathcal{M})$ for a measurable space $(X, \mathcal{M})$, the identity map $(X, \mathcal{A}) \rightarrow (X, \cB)$ is measurable. Similarly, if $\mathcal{A} \sqsubseteq \cB$ for $\mathcal{A}, \cB \in \mathrm{Sub}^{\mathrm{eq}}_\sigma(\mathcal{M})$ then the measurable quotient map $X \rightarrow X /\! \sim_{\cB}$ factors through the quotient $X \rightarrow X /\! \sim_{\mathcal{A}}$.

We start our investigation by looking at the atoms in the lattice $(\mathrm{Sub}_\sigma(\mathcal{M}), \sqsubseteq)$ for some fixed measurable space $(X, \mathcal{M})$. These atoms have been identified in \S 18 of \cite{raorao}, and we describe their construction in the following. The basic idea is to lift equivalence relations on a measurable space $(X, \mathcal{M})$ to the set of probability measures $\cS(\mathcal{M})$. Recall that a $0-1$-measure on $\mathcal{M}$ is a probability measure which only takes the values $0$ and $1$. For instance every point measure $\delta_x$, defined by $\delta_x (A) = 1$ if $x \in A$ and $\delta_x (A) = 0$ otherwise, is a $0-1$-measure.

\begin{prop}
For any family $\{ S_i \}_{i \in I}$ of pairwise disjoint sets of $0-1$-measures the set of $\{S_i\}_{i \in I}$-invariant measurable sets, defined as:
$$\mathcal{M}^{\langle S_i\rangle_{i \in I}} := \{ A \in \mathcal{M} |\: \forall i \in I.\forall \mu, \mu' \in S_i. \: \mu (A) = \mu'(A) \}$$
is a sub-$\sigma$-algebra of $\mathcal{M}$.\label{refx1}
\end{prop}

\begin{prove}
Straightforward using the $\pi-\lambda$-Theorem \cite{eed1}.\qed
\end{prove}

Rao and Rao \cite{raorao} have shown that the atoms in $(\mathrm{Sub}_\sigma(\mathcal{M}), \sqsubseteq)$ are precisely the sub-$\sigma$-algebras of the form $\mathcal{M}^S$ for set $S$ containing two distinct $0-1$-measures.

\begin{lemm}
Let $(X, \mathcal{M})$ be a measurable space and $\mu,\mu'$ be a pair of distinct $0-1$-measures. Then $\mathcal{M}^{\{\mu,\mu'\}}$ is an atom in $(\mathrm{Sub}_\sigma(\mathcal{M}), \sqsubseteq)$. Conversely every atom of $(\mathrm{Sub}_\sigma(\mathcal{M}), \sqsubseteq)$ is of the form $\mathcal{M}^{\{\mu,\mu'\}}$.\label{x5}
\end{lemm}

So let us take a closer look at $0-1$-measures.

\begin{prop}
Let $\mu$ be a $0-1$-measure on a measurable space $(X, \mathcal{M})$. If the set $\bigcap \{A \in \mathcal{M} |\: \mu(A) =1\}$ is nonempty, then it cannot be separated by $\mathcal{M}$. In particular if $(X, \mathcal{M})$ is separated then a $0-1$-measure $\mu$ is either a point measure or $\bigcap \{A \in \mathcal{M} |\: \mu(A) =1\} = \emptyset$.
\end{prop}

\begin{prove}
Suppose otherwise $x,y \in \bigcap \{A \in \mathcal{M} |\: \mu(A) =1\}$ could be separated by $B$, i.e. $x \in B$ and $y \notin B$. Then we get $1 = \mu (X) = \mu(B) + \mu(X \setminus B) = 0$ which yields an obvious contradiction.\qed
\end{prove}

Notice that it is possible for the intersection to be empty. For instance in $(\mathbb{R}, \cC)$, $\cC$ being the countable-cocountable $\sigma$-algebra on $\mathbb{R}$, there is a $0-1$-measure which gives all cocountable sets measure $1$. This counterexample appears amongst others in \cite{raorao}. However, if we restrict ourselves to countably-generated measurable spaces we get the following result which is essentially shown in \S 4 of \emph{op.cit.}.

\begin{lemm}
If $(X,\mathcal{M})$ be a countably-generated measurable space. Then any $0-1$-measure $\mu$ is a point measure.\label{refx2}
\end{lemm}

\begin{coro}
If $(X,\mathcal{M})$ is a separable measurable space then any $0-1$-measure $\mu$ is a unique point measure.\label{ref6b}
\end{coro}

The case of the countable-cocountable $\sigma$-algebra $\cC$ on $\mathbb{R}$ shows that these results do not transfer to sub-$\sigma$-algebras of separable measurable spaces. It also shows another thing, namely that in general the lattices $(\Sub (\mathcal{M}), \sqsubseteq)$ are not atomic, because as a sub-$\sigma$-algebra of the standard Borel sets $\cB$, $\cC$ is not contained in any $\cB^{\{\mu,\mu'\}}$ for a pair of $0-1$-measures.

Surprisingly, this is different in the lattice $(\mathrm{Sub}^{\mathrm{eq}}_\sigma(\mathcal{M}), \sqsubseteq)$.

\begin{prop}
For every measurable space $(X,\mathcal{M})$, the lattice $(\mathrm{Sub}^{\mathrm{eq}}_\sigma (\mathcal{M}), \sqsubseteq)$ is atomic.
\end{prop}

\begin{prsk}
In terms of equivalence relations, the atoms of $(\mathrm{Sub}^{\mathrm{eq}}_\sigma (\mathcal{M}), \sqsubseteq)$ correspond exactly to the relations $\sim^{\{a,b\}}$ which identify elements $a,b \in X$ which can be separated by $\mathcal{M}$. It is not hard to see that every equivalence relation induced by a sub-$\sigma$-algebra of $\mathcal{M}$ is obtained as the supremum of such relations below it, and that this carries over through the Galois connection.\qed
\end{prsk}

Let us now turn towards smooth equivalence relations and countably-generated $\sigma$-algebras. In the following we define the notion of finitely generated partitions on $(X, \mathcal{M})$, show that these are closed under suprema in $(\mathrm{Eq}(X), \lessdot)$ and that they are smooth relations in the case of analytic spaces. We then use these results to show that in general $(\mathrm{Sub}^{\omega}_\sigma(\mathcal{M}), \sqsubseteq)$ is not a dcpo and that the way-below order on it may be trivial.

\begin{defi}
Let $(X, \mathcal{M})$ be a measurable space, and $\{S_f\}_{f \in F}$ be a finite family of pairwise disjoint, finite subsets of $X$. We define a relation $\sim^{\langle S_f\rangle_{f \in F}}$ as the equivalence relation on $X$ which has the equivalence classes $S_f$ for $f \in F$ and $\{x\}$ for $x \in X \setminus \bigcup_{f \in F} S_f$, i.e. $\sim^{\langle S_f\rangle_{f \in F}}$ identifies the elements of the $S_f$. We call such an equivalence relation a \emph{finitely generated partition} of $X$.

For a finitely generated partition  $\{S_f\}_{f \in F}$ of $X$, we call the sub-$\sigma$-algebra of $\sim^{\langle S_f\rangle_{f \in F}}$-invariant subsets of $\mathcal{M}$ an \emph{fgp-sub-$\sigma$-algebra} of $\mathcal{M}$.
\end{defi}

Of course, the fgp-sub-$\sigma$-algebras of $\mathcal{M}$ are precisely the images of the finitely-generated partitions under the Galois connection $(\Sub (\mathcal{M}),\sqsubseteq) \leftrightarrow (\mathrm{Eq}(X), \lessdot)$. In fact, they correspond to the finite suprema of the atoms of $\mathrm{Sub}^{\mathrm{eq}}_\sigma(\mathcal{M})$, as the following result shows whose proof is straightforward.

% Notice furthermore that the finitely generated partitions can be lifted to partitions on the set of $0-1$-measures by identifying the corresponding point measures. Hence, by Lemma \ref{refx2}, in a countably-generated measurable space every atom of $\mathrm{Sub}_\sigma(\mathcal{M})$ is an fgp-sub-$\sigma$-algebra corresponding to a finitely generated partition given by a single set $S$ with two distinct points. Moreover, in this countably-generated case, the sub-$\sigma$-algebras used in Proposition \ref{refx1} are precisely the fgp-sub-$\sigma$-algebras.

\begin{lemm}
Let $X$ be a set. In $(\mathrm{Eq}(X),\lessdot)$ a finite join of finitely-generated partitions is again a finitely generated partition.\label{ref6}
\end{lemm}

\begin{rem}
In fact from the domain-theoretical viewpoint the finitely-generated partitions are of particular interest in $(\mathrm{Eq}(X), \lessdot)$, because the are exactly the compact elements making this lattice algebraic.
\end{rem}

The following result is now a straightforward consequence of the considerations we just made.

\begin{coro}
In a measurable space $(X, \mathcal{M})$, every element of $(\mathrm{Sub}^{\mathrm{eq}}_\sigma (\mathcal{M}), \sqsubseteq)$ is the directed supremum of the fgp-sub-$\sigma$-algebras of $\mathcal{M}$ that are below it.
\end{coro}

As we have seen in section 2.2, for an analytic space $(X, \mathcal{M})$ the set $\mathrm{Sub}^{\omega}_\sigma(\mathcal{M})$ is a subset of $\mathrm{Sub}^{\mathrm{eq}}_\sigma(\mathcal{M})$. We now show that it includes all fgp-sub-$\sigma$-algebras of $\mathcal{M}$.

\begin{prop}
Let $(X, \mathcal{M})$ be an analytic space. Then any fgp-sub-$\sigma$-algebra $\mathcal{A}$ of $\mathcal{M}$ is countably-generated. Equivalently, finitely generated partitions on $X$ are smooth equivalence relations on $(X, \mathcal{M})$.\label{ref7}
\end{prop}

\begin{prsk}
One shows that a finitely-generated partition $\sim^{\langle S_f\rangle_{f \in F}}$ is smooth on $(X, \mathcal{M})$ by applying Theorem \ref{x3} to the kernel of a measurable map which picks one representative for each equivalence class.\qed
\end{prsk}

So let us summarise.

\begin{coro}
For every analytic space $(X, \mathcal{M})$ the poset $(\mathrm{Sub}^{\omega}_\sigma(\mathcal{M}), \sqsubseteq)$ is atomic, i.e. every element can be obtained as the supremum of the atoms below it. Moreover, $(\mathrm{Sub}^{\omega}_\sigma(\mathcal{M}), \sqsubseteq)$ contains all the atoms of $(\mathrm{Sub}_\sigma(\mathcal{M}), \sqsubseteq)$ (and $(\mathrm{Sub}^{\mathrm{eq}}_\sigma(\mathcal{M}), \sqsubseteq)$) and every element of $\mathrm{Sub}^{\mathrm{eq}}_\sigma(\mathcal{M})$ can be obtained as directed supremum of elements of $\mathrm{Sub}^{\omega}_\sigma(\mathcal{M})$. In particular, every smooth equivalence relation on $(X, \mathcal{M})$ is the directed join of the finitely generated partitions included in it.
\end{coro}

With the counterexample of section 3.2 in \cite{eedconglatt} we can show again:

\begin{prop}
For an analytic space $\mathrm{Sub}^{\omega}_\sigma(\mathcal{M})$ is not directed complete in general.\label{refx5}
\end{prop}

Furthermore we can show again that for the usual Borel-$\sigma$-algebra $\cB$ on $\mathbb{R}$ the way-below relation on $(\mathrm{Sub}^{\mathrm{eq}}_\sigma(\cB), \sqsubseteq)$ and $(\mathrm{Sub}^{\omega}_\sigma(\cB), \sqsubseteq)$ is trivial.

\begin{theo}
On $(\mathrm{Sub}^{\mathrm{eq}}_\sigma(\cB), \sqsubseteq)$ and $(\mathrm{Sub}^{\omega}_\sigma(\cB), \sqsubseteq)$ the way-below relation $\ll$ is trivial, i.e. $\mathcal{A} \ll \mathcal{A}'$ if and only if $\mathcal{A} = \cB$.\label{refx4}
\end{theo}

\begin{prsk}
One assumes $\mathcal{A} \neq \cB$ so that there exist distinct $x,y \in \mathbb{R}$ with $x \sim_\mathcal{A} y$. Then one picks a non-$\cB$-measurable set $T$ which separates $x$ and $y$. The finite subsets of $T$ and $\mathbb{R} \setminus T$ induce a directed set of finitely generated partitions whose supremum has as invariant sets only $T$ and $\mathbb{R} \setminus T$. Since these are non-$\cB$-measurable, the corresponding supremum of fgp-sub-$\sigma$-algebras is $\cB$. But all the partitions separate $x$ and $y$ hence the corresponding $\sigma$-algebras cannot be below $\mathcal{A}$.\qed
\end{prsk}

\begin{coro}
The way-below relation on $(\mathrm{Sm}(\mathbb{R}, \cB), \lessdot)$ is trivial.
\end{coro}

This is an unsatisfactory result, especially considering the fact that for any set $X$ the lattice $(\mathrm{Eq}(\mathbb{R}), \lessdot)$ of all equivalence relations of $X$ is algebraic, and thus from a domain-theoretical viewpoint carries a very well-behaved ordere structure.

\section{Conclusion}

For a given measurable space $(X, \mathcal{M})$ we have investigated the properties of the inclusion order and its opposite on the sets $\mathrm{Sub}_\sigma(\mathcal{M})$ (of sub-$\sigma$-algebras of $\mathcal{M}$), $\mathrm{Sub}^{\mathrm{eq}}_\sigma (\mathcal{M})$ (of equivalence-induced sub-$\sigma$-algebras of $\mathcal{M}$) and $\mathrm{Sub}^{\omega}_\sigma (\mathcal{M})$ (of countably-generated sub-$\sigma$-algebras of $\mathcal{M}$) from a domain-theoretical viewpoint. In particular we have shown that in the important case of the standard Borel-structure on the real numbers the way-below relation on these posets is trivial. Along the way we have also proved that the category of analytic spaces does not have all coequalizers, generalising Doberkat's result in \cite{eedconglatt} that it is not closed under pushouts in the category of measurable spaces.

For the inclusion order our results also hold for the $\w$-chain way-below relation. We do not know if this is the case for the opposite order investigated in section 4, as Theorem \ref{refx4} uses a directed set that cannot be reduced to an $\w$-chain. In fact, we also do not know whether for an analytic space an analogous result of Corollary \ref{refx5} holds for $\w$-chains. Such a result would be interesting as the nature of $\sigma$-algebras and smooth relations is inherently related to countable structures and thus $\w$-accessibility might provide a more successful approach than directed accessibility in this case.

\subsection*{Acknowledgments}

The author would like to thank Ernst-Erich Doberkat for his encouragement in doing this research. Helpful discussions with Christoph Schubert were appreciated.

\bibliography{myrefs}
\bibliographystyle{abbrv}

\end{document}

\bibliography{myrefs}
\bibliographystyle{eptcs}